\documentstyle[12pt]{article}
\def\minus{\mbox{$-$}}
\newcommand{\STRUT}{\rule{0in}{2.5ex}}
\newskip\humongous \humongous=0pt plus 1000pt minus 1000pt
\begin{document}
\vspace*{-0.6in}
\hfill \fbox{\parbox[t]{1.12in}{LA-UR-03-3119}}\hspace*{0.35in}
\vspace*{0.0in}

\begin{center}

{\Large {\bf Configuration-Space-Faddeev Born Approximations}}\\

\vspace*{0.4in}
{\bf J.\ L.\ Friar} \\
{\it Theoretical Division,
Los Alamos National Laboratory \\
Los Alamos, NM  87545} \\
\vspace*{0.10in}
\vspace*{0.10in}
{\bf G.\ L.\ Payne}\\
{\it Department of Physics and Astronomy,
University of Iowa\\
Iowa City, IA 52242}\\
\end{center}

\vspace*{0.10in}
\begin{abstract}
Alternative definitions of the Born approximation and the distorted-wave Born
approximation within the framework of the configuration-space Faddeev equations
are explored.  The most natural definition does not correspond to the Born
approximation derived from the Schr\"{o}dinger equation, even though the exact
T-matrices for both formalisms are equivalent.  The Schr\"{o}dinger form is
optimal, although it is shown that the differences are numerically unimportant. 
The DWBA corresponding to the Faddeev equations is not channel symmetric,
although numerically this is unimportant for the p-d (Coulomb) case. Convergence
of the Born approximation partial-wave series is briefly investigated for p-d
and n-d scattering below breakup threshold.
\end{abstract}

\pagebreak

\section{Introduction}

Modern few-nucleon calculations are in principle a rich source of information
about the nuclear force, provided that numerical results can be obtained with
sufficient precision.  The required level of precision depends sensitively on
the observable being calculated.  That level ranges from modest in the case of
unpolarized observables such as the differential cross section to high in the
case of some polarization observables, since the latter often depend on delicate
cancellations between partial waves.  One of the strengths of the few-nucleon
field is that accurate calculations for realistic potentials are now
routine\cite{nd-bench1,nd-bench2, pd-bench}.

One of the consequences of this numerical precision requirement is that a
substantial number of partial waves may be required.  This is not such a dire
problem for momentum-space techniques, many of which generate a Born series for
the T-matrix as the preferred method of solution. Such a series should converge
faster as the angular momentum increases. That is not the case, however, in
configuration space, where each partial wave is solved in its entirety.  Because
more and more partial waves of the potential contribute as the angular momentum
increases, this significantly increasing effort is rewarded by decreasing return
as the size of higher-partial-wave T-matrix elements becomes smaller and
smaller.  One way to increase efficiency in the process is to use the Born
approximation for higher partial waves, which significantly reduces the
effort\cite{kohn}.

We detail below several pitfalls that arise in defining the Born approximation
using configuration-space Faddeev techniques\cite{pd-methods}. While none of
these difficulties are very important in practical calculations (although this
is not {\it a priori} obvious), it is nevertheless worthwhile to avoid them
altogether.  These problems are derived in Section 2 for the Born approximation
(BA) and in Section 3 for the distorted-wave Born approximation (DWBA), and a
few numerical results are given and discussed in Section 4, which is followed by
our summary and conclusions.  We restrict ourselves to energies below the
deuteron-breakup threshold, and we ignore three-nucleon forces, which are an
inessential complication.

\section{Faddeev Equations and the Born Approximation}

The Faddeev decomposition in configuration space\cite{noyes} can be pictured
most easily as dividing the wave function into parts corresponding to different
asymptotic processes.  For three (identical) particles this leads to three
equations
$$
(E - T -V_i) \psi_i = V_i (\psi_j + \psi_k)\, , \eqno(1)
$$
where $i,j,k$ take on the values of the cyclic permutations of 1, 2, and 3, and
$V_i \equiv V({\bf x}_i)$. The Schr\"{o}dinger wave function is given by
$$
\Psi = \psi_1({\bf x}_1,{\bf y}_1) + \psi_2({\bf x}_2,{\bf y}_2) +
       \psi_3({\bf x}_3,{\bf y}_3)\,, \eqno(2)
$$
with arguments fashioned from the coordinates, ${\bf r}_i$, of all $i$ 
particles:
$$
{\bf x}_i  \equiv  {\bf r}_j - {\bf r}_k\,, \eqno(3a)
$$
and
$$
{\bf y}_i \equiv  \frac{1}{2} ({\bf r}_j + {\bf r}_k) - {\bf r}_i\, .
\eqno(3b)
$$
The coordinate ${\bf x}_i$ is the distance between particles $j$ and $k$, while
${\bf y}_i$ is the distance between particle $i$ and the center of cluster $j +
k$. Depending on the desired application a normalization factor of $1 /
\sqrt{3}$ may be used in the definition of $\Psi$\cite{review}.  With these
definitions the sum of the three independent Eqns.~(1) yields the original
Schr\"{o}dinger equation
$$
(E - T - V_1 - V_2 - V_3) \Psi = 0\, . \eqno(4)
$$
The coordinates (3) used in the Faddeev wave functions in Eqn.~(4) generate
different asymptotic configurations (or channels) that facilitate the
appropriate boundary conditions for this partial-differential equation.  Used in
this way the Faddeev equations are merely another (equivalent) form of the
original Schr\"{o}dinger equation.

The appropriate boundary conditions for elastic scattering (below breakup
threshold) are easily seen from Eqn.~(1).  For large $|{\bf y}_i|$ the scattered
wave function factorizes into the product of a deuteron wave function $\phi_{\rm
d} ({\bf x}_i)$ and an outgoing wave, $\chi ({\bf y}_i)$, which guarantees that
the right-hand side of Eqn.~(1) vanishes in this limit.  This product of
$\phi_{\rm d}$ and  a  plane wave (or a Coulomb wave in case of p-d scattering),
$\chi ({\bf y}_i)$, defines the ``free'' channel wave function,
$\phi_i$\cite{review}:
$$
\phi_i({\bf x}_i,{\bf y}_i) = \phi_d({\bf x}_i) \chi({\bf y}_i)\,, \eqno(5a)
$$
satisfying
$$
(E - T - V_i ) \phi_i = 0 \,, \eqno(5b)
$$
which we can use to construct the exact T-matrix and its Born approximation.

The Faddeev T-matrix is most easily developed by projecting Eqn.~(1) with
$\langle \phi_i |$.  The resulting left-hand-side integral vanishes because of
Eqn.~(5b), except for surface terms that constitute the T-matrix (just like the
two-body case\cite{schiff})
$$
t_{\rm F} = \langle \phi_1 | V_1 | \psi_2 + \psi_3 \rangle \,, \eqno(6)
$$
and any other permutation of Eqn.~(1) yields the same result.  This can be
converted\cite{walter} to a more recognizable form by rearranging Eqn.~(5b) to
$\langle \phi_1 | V_1 = \langle \phi_1 | (E - T)$, making use of the Hermiticity
of the Hamiltonian for these wave functions below breakup threshold, and
employing Eqn.~(1) to produce the Schr\"{o}dinger form of the
T-matrix\cite{satchler}
$$
t_{\rm S} = \langle \phi_1 | V_2 + V_3 | \Psi \rangle \, . \eqno(7)
$$
The potential $V_2 + V_3$ is the usual ``inter-cluster'' 
potential\cite{satchler} between the nucleon projectile and the deuteron target
that generates the scattering T-matrix.  The two forms of the T-matrix given in
Eqns.~(6) and (7) are exactly equivalent if the wave functions are exact.

Born approximations can be obtained from Eqns.~(6) and (7) by replacing the
$\psi_i$ by $\phi_i$  and  $\Psi$ by $\Phi=\phi_1+\phi_2+\phi_3$:
$$
t_{\rm F}^{\rm BA} = \langle \phi_1 | V_1 | \phi_2 + \phi_3 \rangle \,, 
\eqno(8a)
$$
and
$$
t_{\rm S}^{\rm BA} = \langle \phi_1 | V_2 + V_3 | \Phi \rangle \, . \eqno(8b)
$$
In order to facilitate the comparison of the two Born approximations in
Eqns.~(8) it is convenient to introduce the compact notation
$$
(ijk) \equiv \langle \phi_i | V_j | \phi_k \rangle \, . \eqno(9)
$$
Interchanging the labels of pairs of variables leads to (132)$\equiv$(123),
etc., while the trick used to convert Eqn.~(6) to Eqn.~(7) leads to
(122)$\equiv$(112), etc.  We thus arrive at
$$
t_{\rm F}^{\rm BA} = 2 (112) \,, \eqno(10a)
$$
and
$$
t_{\rm S}^{\rm BA} = 2 (112) + 2 (121) + 2 (123) \,, \eqno(10b)
$$
which are distinctly different.  One can also use these tricks to demonstrate
that the Faddeev form of the Born-approximation T-matrix is channel symmetric:
$$
t^{\rm ab}_{\rm F} = \langle \phi^{\rm a}_1 | V_1 | \phi^{\rm b}_2 +
\phi^{\rm b}_3 \rangle = \langle \phi^{\rm a}_2 + \phi^{\rm a}_3 | V_1 |
\phi_1^{\rm b} \rangle = t^{\rm ba}_{\rm F} \, , \eqno(10c)
$$
where we have attached channel labels that were previously suppressed. In other
words the ``post'' and``prior'' forms of the Born approximation are identical, a
result long known for $t_{\rm S}$\cite{satchler}.  The numerical significance of
the difference of Eqns.~(10a) and (10b) will be discussed in Section~(4).

An important point concerning the Born approximation has been made by
Kievsky {\it et al}\cite{kohn}. The Kohn variational functional can be written
as\cite{review}
$$
I(\Psi) = \langle \Psi | H - E | \Psi \rangle = \langle \psi_1 | H - E | 
\Psi \rangle  \,, \eqno(11)
$$
where the last step follows by interchange of variables and by removing the
normalization factor $1 / \sqrt{3}$ from each wave function.  Varying $\Psi$
(and $\psi_1$) about the {\bf plane-wave} limit leads to
$$
t_{\rm Kohn} = \langle \phi_1 | V_2 + V_3 | \Phi \rangle + {\cal{O}}( V^2 )\,, 
\eqno(12)
$$
since any change in $\Psi$ beyond the plane-wave part must be proportional to
the potential, and the Kohn error is proportional to the square of that change
in $\Psi$.  For this reason $t^{\rm BA}_{\rm S}$ in Eqn.~(8b) gives the optimal
Born approximation.  The two terms missing in $t^{\rm BA}_{\rm F}$ are
``hidden'' in $\psi_2 + \psi_3$.  Even though $t^{\rm BA}_{\rm F}$ is linear in
the potential $V_1$, that potential serves partly to drive the dynamics (since
Eqn.~(6) is exact) and partly to bind the deuteron; the latter terms do not
directly contribute to the T-matrix, and serve to ``hide'' $V_2$ and $V_3$ terms
residing in $\psi_2$ and $\psi_3$. Thus the Faddeev Born approximation has
first-order (in $V$) errors, unlike the Schr\"{o}dinger result.

\section{Faddeev Equations and the DWBA}

The remaining task is to compare distorted-wave Born approximations from the
Faddeev and the Schr\"{o}dinger equations.  This is complicated by the fact that
long-range forces (such as Coulomb) have been treated in the Faddeev formalism
in a variety of ways, reflecting the non-unique character of the Faddeev
equations themselves.  A powerful formal device is the introduction of Faddeev
distortion potentials, $X_i$.  We can replace Eqn.~(1) by\cite{review}
$$
[E - T -V_i -X_j - X_k] \psi_i = (V_i - X_i) (\psi_j + \psi_k) \, . \eqno(13)
$$
If one sums the three independent (cyclic-permutation) members of this set, one
obtains the original Schr\"odinger equation in Eqn.~(4) (i.e., the $X_i$
cancel), and $\Psi$ is free of the arbitrary $X_i$, as well, although individual
$\psi_i$ are not free. It is effective to choose the  $X_i$ so that long-range
distortions are built into the wave functions $\psi_i$.  As a specific example
we treat the important case of long-range Coulomb distortions, and write $V_i =
V_i^{\rm st} + V_i^{\rm C}$, where $V_i^{\rm st}$ and $V_i^{\rm C}$ are the
strong-interaction and Coulomb parts of the nuclear force.  In this case we have
found it most efficient numerically to choose $X_i = V_i^{\rm C}$, which places
all of the Coulomb interaction on the left-hand side of Eqn.~(13).  Other
choices are possible and have been used by others\cite{s-s,unique}.

We begin our treatment of DWBA by noting that if a long-range force modifies the
asymptotic form of the inter-cluster scattering wave function, $\chi ({\bf
y}_i)$, our basis wave functions (i.e., the ``free'' scattering wave functions)
are modified and correspond to
$$
(E - T - V_1 - U_1) \phi_1 = 0 \,, \eqno(14)
$$
where $U_1$ is an appropriate long-range interaction potential (e.g., $Z
\alpha/y$  for p-d Coulomb scattering).  We next rearrange Eqn.~(13) (choosing
$i=1$) by moving all $X$-terms to the right-hand side, and then adding and
subtracting a potential $U_1$:
$$
(E - T - V_1 - U_1) \psi_1 = (X_2 + X_3 - U_1) \psi_1 + (V_1 - X_1) 
(\psi_2 + \psi_3) \, . \eqno(15)
$$
Projecting this equation on the left with $\langle \phi_1 |$ gives a
left-hand-side that vanishes because of Eqn.~(14) except for surface terms (that
comprise the T-matrix, as before)
$$
\bar{t}_{\rm F} = \langle \phi_1 | X_2 + X_3 - U_1 | \psi_1 \rangle + 
\langle \phi_1 | V_1 - X_1 | \psi_2 + \psi_3 \rangle \,, \eqno(16)
$$
where we ignore here (and hereafter) the T-matrix coming solely from the
potential $U_1$ (e.g., the point-Coulomb T-matrix), which is important but
peripheral to our discussion. The barred-t indicates asymptotic states modified
by $U_1$. Choosing  $X_i = V_i^{\rm C}$ in the Coulomb case leads to our
preferred form of the Coulomb-distorted-wave Faddeev T-matrix, $\bar{t}_{\rm
F}^{\, \rm C}$:
$$
\bar{t}_{\rm F}^{\, \rm C} = \langle \phi_1 | V_1^{\rm st} | \psi_2 + \psi_3 
\rangle + \langle \phi_1 | V^{\rm C} - U_1 | \psi_1 \rangle \,, \eqno(17)
$$
where $V^{\rm C}$ is $V^{\rm C}_2 + V^{\rm C}_3$, and we have used the fact that
$V^{\rm C}_1$ vanishes. If we again use $\langle \phi_1 | V_1 = \langle \phi_1 |
(E - T)$ in Eqn.~(16), we obtain after some algebra
$$
\bar{t}_{\rm S} = \langle \phi_1 | V_2 + V_3 - U_1 | \Psi \rangle \,, \eqno(18)
$$
a remarkably different-looking, but equivalent, result.  Equation~(18) is the
usual Schr\"{o}dinger form of the distorted-wave T-matrix, and differs from
Eqn.~(7) only by the $U_1$ term\cite{satchler}.

Although the two forms of the T-matrix in Eqns.~(16) and (18) are formally
identical, they lead to very different DWBAs. Replacing $\psi_i$ by $\phi_i$ and
$\Psi$ by $\Phi$ in Eqn.~(16) leads to a formally unacceptable DWBA, because it
would depend on the unphysical (i.e., artificial) $X_i$
$$
\bar{t}_{\rm F}^{\rm \, DWBA} = \langle \phi_1 | X_2 + X_3 - U_1 | \phi_1 
\rangle + \langle \phi_1 | V_1 - X_1 | \phi_2 + \phi_3 \rangle \,, \eqno(19)
$$
although the {\bf choice} of $X_i$ leading to Eqn.~(17) obscures this fact. The
DWBA obtained from Eqn.~(18)
$$
\bar{t}_{\rm S}^{\rm \, DWBA} = \langle \phi_1 | V_2 + V_3 - U_1 | \Phi \rangle 
\,, \eqno(20)
$$
is free of that disease because Eqn.~(18) depends only on $\Psi$, and not the
individual $\psi_i$. The Schr\"odinger DWBA is the optimal one in the sense that
corrections to it are second order in the potential. The techniques that led to 
Eqn.~(10c) also demonstrate that the distorted-wave T-matrix in Eqn.~(19) is not
symmetric under interchange of incoming and outgoing channels (post and prior 
forms), because of terms involving $X_1$ and $U_1$. This is a serious disease,
since it violates a symmetry, although we shall see in the next section that the
effect is sufficiently small that it is not a significant practical problem.
Equation~(20) can be shown by the same techniques to be channel symmetric.

In summary the ``Faddeev'' form of the DWBA has three formally serious problems:
(1) it leads to a non-symmetric T-matrix; (2) it leads to manifestly non-unique
results (viz., the $X$-terms); (3) it leads to errors linear in the potential. 
How important these flaws are in a numerical sense is discussed next.

\section{Numerical Results}

We briefly report two numerical results on:  (1) differences between the Faddeev
and Schr\"{o}dinger forms of the Born approximation, together with the magnitude
of the lack of symmetry of the post and prior forms of the T-matrix; (2) the
fractional errors in the difference of the exact Faddeev and Born approximation
results for both p-d and n-d scattering below deuteron-breakup threshold.

The form of the Born approximations given by Eqns.~(10a) and (10b) suggests that
numerical differences should be small.  Below the threshold for deuteron breakup
the region of configuration space where scattering can occur is very limited. 
The deuteron bound-state wave function has a finite extent, $R_d$, while the
nuclear force has a significantly smaller range, $R_V$.  For this reason the
integrand of the configuration (123), for example, is important only in the very
small region where the nuclear potential in variable ${\bf x}_2$  overlaps with
a deuteron wave function in variable ${\bf x}_1$ and another in variable ${\bf
x}_3$.  This occurs only for very small $|{\bf y}_1|$, where the wave function 
is suppressed for large values of the angular momentum.  By far the largest
configuration will be (112), since $V_1$ overlaps easily with $\phi_{\rm d}({\bf
x}_1)$, and this result has a much larger overlap with the deuteron wave
function in variable ${\bf x}_2$.  Numerical results bear this out; the (112)
matrix element completely dominates.  This immediately suggests that the lack of
symmetry between (off-diagonal) T-matrix elements is also numerically small.

\begin{table}[t]
\begin{center}
{\bf Table I}.  Maximum absolute values of the fractional differences between
exact and comparison calculations for differential cross sections
and analyzing powers versus partial wave. The comparison labelled Truncation
deletes the labelled partial wave and all higher ones. The comparison labelled
Born Substitution substitutes the Born approximation for the exact T-matrix
elements for the labelled partial wave and all higher ones.  Results are
presented in the format x.x[n] $\equiv$ x.x\,10$^n$.\\

\vspace*{0.25in}

\begin{tabular}{|l|rrrrrr|}
\hline
\multicolumn{1}{|c}{\STRUT Partial Wave}&
\multicolumn{1}{c}{$5\over2$}&
\multicolumn{1}{c}{$7\over2$}&
\multicolumn{1}{c}{$9\over2$}&
\multicolumn{1}{c}{$11\over2$}&
\multicolumn{1}{c}{$13\over2$}&
\multicolumn{1}{c|}{$15\over2$}\\ \hline \hline
\multicolumn{7}{|c|}{\STRUT Truncation $\minus$ 3 MeV p-d}\\ \hline
$\STRUT \Delta(d\sigma/d\Omega)/(d\sigma/d\Omega)$  &{   }  4.9[-1] &{   }  
1.1[-1] &  {   } 2.8[-2] &{   } 7.8[-3] &{   } 2.0[-3] &{   } 3.9[-4] \\
$\Delta(A_y)/A_y$ &  3.1[{ }1] &  4.4[{ }0] &  1.5[{ }0] &  3.9[-1] 
&{   } 8.6[-2] & {   }1.8[-2]\\ \hline
\end{tabular}

\vspace*{0.25in}

\begin{tabular}{|l|rrrr|}
\hline
\multicolumn{1}{|c}{\STRUT Partial Wave}&
\multicolumn{1}{c}{$5\over2$}&
\multicolumn{1}{c}{$7\over2$}&
\multicolumn{1}{c}{$9\over2$}&
\multicolumn{1}{c|}{$11\over2$}\\ \hline \hline
\multicolumn{5}{|c|}{\STRUT Born Substitution $\minus$ 3 MeV n-d}\\ \hline
$\STRUT \Delta(d\sigma/d\Omega)/(d\sigma/d\Omega)$  &{   }  1.4[-1] 
&{   }  2.2[-3] &  {   } 1.2[-4] &{   } 8.9[-6] \\
$\Delta(A_y)/A_y$ &  4.1[{ }0] &  2.2[-1] &  1.9[-2] &  1.9[-3] \\ \hline
\multicolumn{5}{|c|}{\STRUT Born Substitution $\minus$ 1 MeV p-d}\\ \hline
$\STRUT \Delta(d\sigma/d\Omega)/(d\sigma/d\Omega)$  &  7.0[-2] &  7.1[-4] 
&  1.3[-5] &  4.3[-6] \\
$\Delta(A_y)/A_y$ &  8.6[{ }0] &  1.3[-1] &  1.1[-2] &  2.7[-3] \\ \hline
\multicolumn{5}{|c|}{\STRUT Born Substitution $\minus$ 3 MeV p-d}\\ \hline
$\STRUT \Delta(d\sigma/d\Omega)/(d\sigma/d\Omega)$  &{   }  8.5[-2] 
&{   }  2.1[-3] &  {   } 9.7[-5] &{   } 5.8[-6] \\
$\Delta(A_y)/A_y$ &  4.2[{ }0] &  1.4[-1] &  1.3[-2] &  1.1[-3] \\ \hline
\end{tabular}

\end{center}
\end{table}

The importance of the Born amplitudes for higher partial waves (in $J$, the
total angular momentum of the channel states) is indicated in Table I, where
both the differential cross section, $d \sigma/d \Omega$, and analyzing power,
$A_y$, have been computed using the AV18 potential\cite{AV18}. These observables
were computed in two ways: exact calculations through $J=11/2$ (supplemented by
Born approximation results for higher waves where needed) and various comparison
calculations described below. The maximum value of the magnitude of the
fractional difference between the exact and comparison calculations was computed
for the angular range $\theta = 30^{\circ} - 150^{\circ}$. The restricted
angular range was used because the analyzing power gets small outside that
range, and relatively large fractional errors in that regime would be
meaningless. In the small table labelled Truncation the comparison calculation
had its partial-wave series truncated at the labelled $J$ (i.e., that $J$-wave
and all higher ones were omitted), leading to some very large errors. If one
adopts the criterion that observables should be calculated to 1\% accuracy, one
needs to keep all matrix elements through $J=9/2$ for the differential cross
section, and through $J=15/2$ for the analyzing power for 3 MeV p-d scattering.

In the tables labelled Born Substitution the comparison case involved
substituting the (Schr\"odinger) Born amplitude for the labelled wave and all
higher ones through the maximum value of $J=15/2$. The differential cross
sections are accurately computed (using the 1\% criterion) by keeping exact
matrix elements through $J=5/2$, while the analyzing power requires exact
$J=9/2$ matrix elements. Exact configuration-space Faddeev calculations for
large values of $J$ (such as 11/2) require significant computational resources,
and should be avoided if possible.

The difference between the Faddeev Born approximation and the Schr\"odinger Born
approximation depends on the particular matrix element, but typically is only a
few per cent, and this is not likely to be very important. The Schr\"odinger
form of the BA and DWBA is nevertheless demonstrably closer to the exact matrix
elements for higher partial waves than is the Faddeev form of those matrix
elements. The channel asymmetry in p-d scattering matrix elements at 3 MeV is
typically a few times $10^{-3}$, and hence is numerically unimportant for the 
Coulomb DWBA problem.

In summary, we have shown that the most natural Born approximation for the
Faddeev T-matrix differs from that of the Schr\"odinger equation. We have shown
that the Faddeev DWBA has three formally serious properties: (1) it leads to a
non-symmetric T-matrix; (2) it leads to manifestly non-unique results; (3) it
leads to errors linear in the potential. In practical (i.e., numerical) terms
these defects are not very important. Our analysis demonstrates that the
dominant part of the Born approximation is contained in both approaches, and
they differ only in much smaller contributions.

\begin{center}
{\large {\bf Acknowledgments}}\\
\end{center}
The work of JLF was performed under the auspices of the U. S. Department of
Energy, while that of GLP was supported in part by the U. S. Department of
Energy.


\begin{thebibliography}{999}

\bibitem{nd-bench1} D. H\"uber, W. Gl\"ockle, J. Golak, H. Wita\l a,
H. Kamada, A. Kievsky,  S. Rosati, and M. Viviani, {\it Phys. Rev. C}
{\bf 51}, 1100 (1995).

\bibitem{nd-bench2} A. Kievsky, M. Viviani, S. Rosati, D. H\"uber, W. Gl\"ockle,
H. Kamada, H. Wita\l a, and J. Golak, {\it Phys. Rev. C} {\bf 58}, 3085 (1998).

\bibitem{pd-bench} A. Kievsky, J. L. Friar, G. L. Payne, S. Rosati, and M. 
Viviani, {\it Phys. Rev. C} {\bf 63}, 064004 (2001).

\bibitem{kohn} A. Kievsky, S. Rosati, W. Tornow, and M. Viviani, 
{\it Nucl. Phys.} {\bf A607}, 402 (1996); C. R. Brune, H. J. Karwowski, 
E. J. Ludwig, K. D. Veal, M. H. Wood, A. Kievsky, S. Rosati, M. Viviani, 
{\it Phys. Lett.} {\bf B406}, 292 (1997).

\bibitem{pd-methods} C. R. Chen, J. L. Friar, G. L. Payne, 
{\it Few-Body Systems} {\bf 31}, 13 (2001).

\bibitem{noyes}  H. P. Noyes, in {\it Three-Body Problem in Nuclear and 
Particle Physics}, ed. by J. S. C. McKee and P. M. Rolph
(North-Holland, Amsterdam, 1970), p. 2.

\bibitem{review}  J. L. Friar and G. L. Payne, in {\it Coulomb Interactions 
in Nuclear and Atomic Few-Body Collisions}, ed. by F. S. Levin and D. A. Micha, 
(Plenum Press, New York, 1996), p. 97.

\bibitem{schiff} L. I. Schiff, {\it Quantum Mechanics}, (McGraw-Hill, 
New York, 1968). 

\bibitem{walter} W. Gl\"ockle, {\it The Quantum Mechanical Few-Body Problem},
(Springer-Verlag, Berlin, 1983).

\bibitem{satchler} G. R. Satchler, {\it Direct Nuclear Reactions}, (Clarendon 
Press, Oxford, 1983).

\bibitem{s-s} T. Sasakawa and T. Sawada, {\it Phys. Rev. C} {\bf 20}, 
1954 (1979).

\bibitem{unique} J. L. Friar, B. F. Gibson, D. R. Lehman, and G. L. Payne, 
{\it Phys. Rev. C} {\bf 37}, 2859 (1988).

\bibitem{AV18} R.B. Wiringa, V.G.J. Stoks, R. Schiavilla, {\it Phys. Rev. C} 
{\bf 51}, 38 (1995).

\end{thebibliography}
\end{document}